\documentclass[twocolumn, prl, superscriptaddress]{revtex4-1}
\usepackage{amssymb}
\usepackage{amsmath}
\usepackage{epsfig}
\usepackage{graphics, graphicx}
\usepackage{bbold}
\usepackage{psfrag}
\usepackage{mathcomp}
\usepackage{subfigure}
\usepackage{verbatim}
\usepackage{color}
\usepackage[colorlinks,citecolor=blue]{hyperref}
\def\cp#1{\mathbf{#1}}

\begin{document}

\title{Visualizing Efimov Correlations in the Bose Polaron}
\author{Mingyuan Sun}
\affiliation{Institute for Advanced Study, Tsinghua University, Beijing, 100084, China}
\author{Hui Zhai}
\affiliation{Institute for Advanced Study, Tsinghua University, Beijing, 100084, China}
\affiliation{Collaborative Innovation Center of Quantum Matter, Beijing, 100084, China}
\author{Xiaoling Cui}
\email{xlcui@iphy.ac.cn}
\affiliation{Beijing National Laboratory for Condensed Matter Physics, Institute of Physics, Chinese Academy of Sciences, Beijing, 100190, China}
\date{\today}

\begin{abstract}
The Bose polaron is a quasi-particle of an impurity dressed by surrounding bosons. 
%Since it is known from the few-body physics that an impurity can form a sequence of Efimov bound states with two bosons on the vicinity of a Feshbach resonance, one would expect that this Efimov correlation can manifest itself in the Bose polaron problem. 
In few-body physics, it is known that two identical bosons and a third distinguishable particle can form a sequence of Efimov bound states in the vicinity of inter-species scattering resonance. On the other hand, in the Bose polaron system with an impurity atom embedded in many bosons, no signature of Efimov physics has been reported in the existing spectroscopy measurements up to date. In this work, we propose that a large mass imbalance between a light impurity and heavy bosons can help %enhance the signature of the Efimov state 
produce visible signatures of Efimov physics in such a spectroscopy measurement. Using the diagrammatic approach in the Virial expansion to include three-body effects from pair-wise interactions, we determine the impurity self-energy and its spectral function. Taking $^{6}$Li-$^{133}$Cs\ system as a concrete example, we find two visible Efimov branches in the polaron spectrum, as well as their hybridizations with the attractive polaron branch. %In this case, the Bose polaron also features narrow spectral widths, suggesting more well defined quasi-particles comparing to the cases with smaller mass imbalance. 
We also discuss the general scenarios for observing the signature of Efimov physics in polaron systems. This work paves the way for experimentally exploring intriguing few-body correlations in a many-body system in the near future. 
\end{abstract}
\maketitle

Top-down and bottom-up are two major approaches to studying correlations in a quantum many-body system. The cold atom system has intrinsic advantage for the bottom-up approach since it is a dilute system and the few-body problems therein are well understood. In this approach, one would like to understand how many-body physics is built up from few-body correlations. In cold atom system, one of the most intriguing three-body correlations lies in Efimov physics, which is characterized by an infinite number of trimer states nearby a two-body resonance and following a universal scaling law \cite{Efimov,Braaten}. Efimov physics has been observed in a number of cold atoms experiments, while all of them are at the few-body level\cite{Efimov_Exp0,Efimov_Exp1,Efimov_Exp1bu,Efimov_Exp3,Efimov_Exp4,Efimov_Exp5,Efimov_Exp9,Efimov_Exp10,Efimov_Exp6,Efimov_Exp7,Efimov_Exp8,Efimov_Exp11,rf_1,rf_2,scaling_1,scaling_2,scaling_3}. The manifestation of Efimov physics in the many-body system has yet to be observed. 

In this context, a convenient and non-trivial testbed is the highly-polarized ultracold gases, which consist of minority impurity atoms interacting with the majority of fermionic or bosonic atoms, respectively called the Fermi or the Bose polarons. Lots of theoretical efforts have been paid to study the Fermi polaron \cite{Chevy, Lobo, Combescot1, Combescot2, Prokofev, Punk, Enss, Cui, Troyer, Bruun, Parish, Zhou, Zinner1, Nishida, Cui1, Cui2} and the Bose polaron \cite{Pitaevskii,Timmermans, Blume, Jaksch, Huang, Schmidt, LiWeiran, Wouters, Demler1, Demler2, Devreese, Zinner2, Levinsen1, Levinsen2, Giorgini, Shchadilova}. Nearby a Feshbach resonance, a Fermi polaron displays an attractive branch \cite{Chevy, Lobo, Combescot1, Combescot2, Prokofev, Punk, Enss} and a repulsive branch \cite{Cui, Troyer, Bruun}, which directly manifests two-body correlations in this system.
% The existence of these two branches are actually hinted by a $1+1$-problem (one impurity plus one majority atom). 
In the past few years, the Fermi polaron has been studied by a number of experiments \cite{Zwierlein,Salomon,Grimm,Kohl,Grimm2016,Roati}, while the Bose polaron has only recently been explored\cite{Aarhus,JILA,Lamb}. 
%the Aarhus \cite{Aarhus} and the JILA \cite{JILA} groups. 
Most of these experiments are the injection radio-frequency spectroscopy measurements, with which both the repulsive and the attractive branches have been observed \cite{Grimm,Kohl,Grimm2016,Roati,Aarhus,JILA}.

From the bottom-up point of view, a difference between the Bose polarons \cite{Aarhus,JILA,Lamb} and the Fermi polarons \cite{Zwierlein,Salomon,Grimm,Kohl,Grimm2016,Roati} already exists in the three-body system consisting of two majority atoms and a third distinguishable particle (usually denoted by "BB+X"),
%$1+2$-problem (one impurity plus two majority atoms), 
where the Bose systems exhibit the Efimov effect while the Fermi systems do not, because Efimov physics is facilitated by the Bose statistics \cite{Efimov, Braaten}. So far, the spectroscopy measurements of the Bose polarons by the Aarhus \cite{Aarhus} and JILA \cite{JILA} groups have not detected such extra Efimov correlation. Despite a few theoretical investigations of the three-body correlations in the Fermi polaron \cite{Parish, Zhou, Zinner1, Nishida, Cui1, Cui2} and the Bose polaron \cite{Levinsen1, Levinsen2, Giorgini}, it is still not clear under what circumstances, the spectroscopy measurement can reveal this difference. Nevertheless, the theoretical treatment of the Bose polaron problem is quite challenging, as it should work for the strong coupling regime and take into account the three-body effects in a non-perturbative way. So far the theoretical tools for this purpose have been quite limited, including the variational approach with truncated number of boson excitations \cite{Levinsen2} and the diffusion Monte Carlo method\cite{Giorgini}. It is thus imperative to develop an alternative method with controllable approximation for the problem in order to further guide the experiments.

\begin{figure}[t]
\includegraphics[width=8.7cm,height=2.9cm]{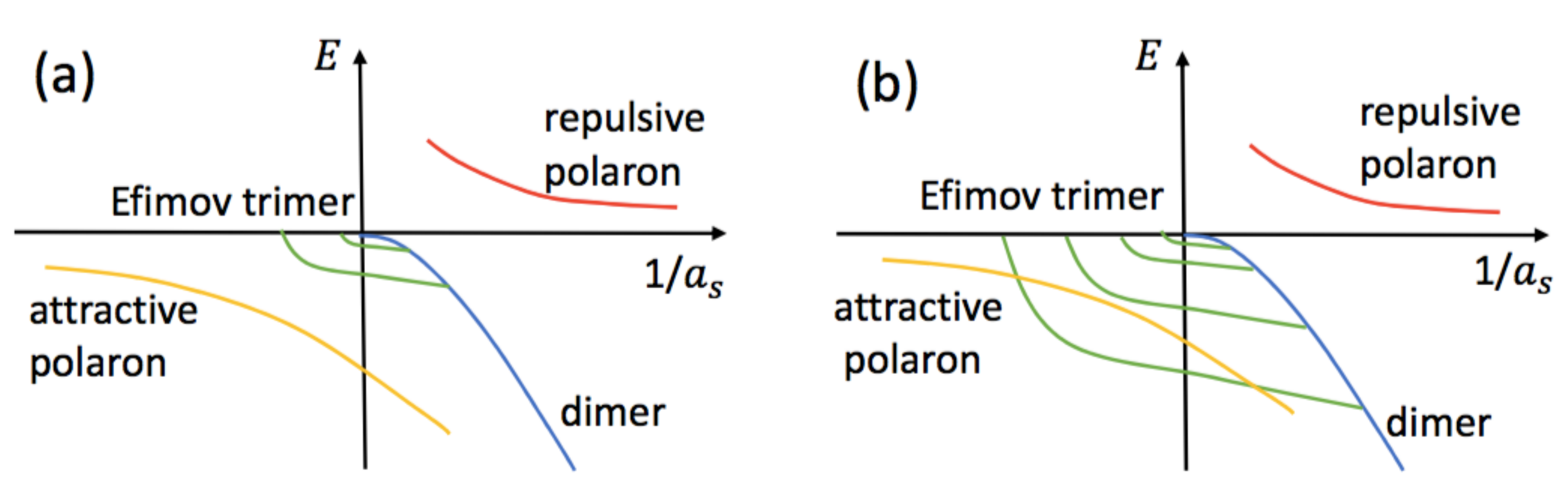}
\caption{(Color Online). Schematics of two scenarios of Efimov trimers in vacuum (green lines) with respect to attractive and repulsive branches of a Bose polaron (orange and red lines). (a) supports shallow Efimov trimers with large scaling factor, while (b) supports reasonably deep trimers with small scaling factor. In (b), the trimer levels can be close or level-crossing with the attractive polaron branch. The signature of Efimov physics will be visible in the spectral function of the Bose polaron in (b) but is hardly visible in (a).
\label{schematic}}
\end{figure}

Before going to details, let us summarize that, with the explicit calculation presented in this work, we can understand the challenge of observing the signature of Efimov physics in the Bose polarons as follows: Comparing the size of Efimov trimers in vacuum $l_t$ and the mean distance of background many-body system $d$, if $l_t\gg d$, which usually occurs for shallow Efimov trimers near resonance (see Fig.\ref{schematic}(a)), their effect can be easily washed out by two-body correlations and is very difficult to resolve in experiments; if $l_t\ll d$, which occurs for deep Efimov trimers, their effect is also difficult to resolve in the injection spectroscopy of polarons due to the little wave function overlap with the initial scattering state. Therefore the most favorable situation is $l_t\sim d$.

In this work, we propose to utilize the large mass imbalance between the impurity and the bosons to facilitate the observation of Efimov correlations in the Bose polarons. Our main results can be illustrated in Fig.\ref{schematic} by comparing two scenarios classified by the mass ratio $\eta=m_\text{b}/m_\text{i}$, where $m_\text{b}\ (m_\text{i})$ is the boson (impurity) mass. For the Efimov trimers of heteronuclear atomic systems, when $\eta\ll 1$, the scaling factor is large\cite{Braaten}, and the trimers are generally quite shallow and appear only close to the resonance\cite{Greene}, see Fig.\ref{schematic}(a).   
%Here we should remark that for the $1+2$-problem considered here, only the interaction between impurity and majority bosons are in the vicinity of resonance, while the interaction between bosons is generically quite weak and is taken as non-interacting in this work for simplicity. In this case, there are two scenarios depending on the mass ratio $\eta=m_\text{b}/m_\text{i}$ ($m_\text{b}$ is the boson mass and $m_\text{i}$ is the impurity mass), as schematically shown in Fig. \ref{schematic}(a) and (b).
%When $\eta\ll 1$, it is generically true that the Efimov trimers are quite shallow and the scaling factor is quite large\cite{Braaten, Greene}, see Fig. \ref{schematic}(a). Thus the Efimov correlation is hardly visible considering $l_t\gg d$. 
Thus the Efimov correlation is hardly visible in the Bose polarons considering $l_t\gg d$. The Aarhus experiment with two different hyperfine states of $^{39}$K ($\eta=1$)\cite{Aarhus} and the JILA experiment with $^{40}$K impurity in ${}^{87}$Rb ($\eta=87/40$)\cite{JILA} both belong to this scenario. %Therefore the Efimov correlation is hardly visible there. 

When $\eta\gg 1$, the scaling factor is small and the Efimov spectrum is dense \cite{Braaten}; meanwhile, the lowest Efimov trimer can appear far from resonance and can be quite deeply bound at resonance\cite{Greene}. Thus, some of the trimers can have the chance to fall into the $l_t\sim d$ regime which makes Efimov signatures visible in the Bose polarons, and the visibility can be further enhanced if these trimers are very close or level-crossing with the attractive polaron branch, as shown in Fig. \ref{schematic}(b). 
%In this scenario, the Efimov correlation can become easily visible in the Bose polaron spectrum. 
Fortunately, taking the experimentally well studied $^6$Li-$^{133}$Cs system as an example, our calculation shows two visible Efimov branches in the spectral response of an $^{6}$Li impurity immersed in $^{133}$Cs bosons, and their hybridizations with the attractive polaron branch causing the spectral broadening and enhanced Efimov signals. 
%Moreover, we find in this case %the Bose polaron possesses narrow spectral widths and  the attractive and repulsive branches are disconnected near the resonance, in contrast to those with smaller mass imbalance as in Aarhus\cite{Aarhus} and JILA\cite{JILA} experiments. 
The unique response properties revealed in this work suggest that the highly mass-imbalanced polaron systems can serve as an ideal platform for detecting intriguing few-body correlations in a many-body environment.    

\begin{figure}[b] 
\includegraphics[width=8.5cm]{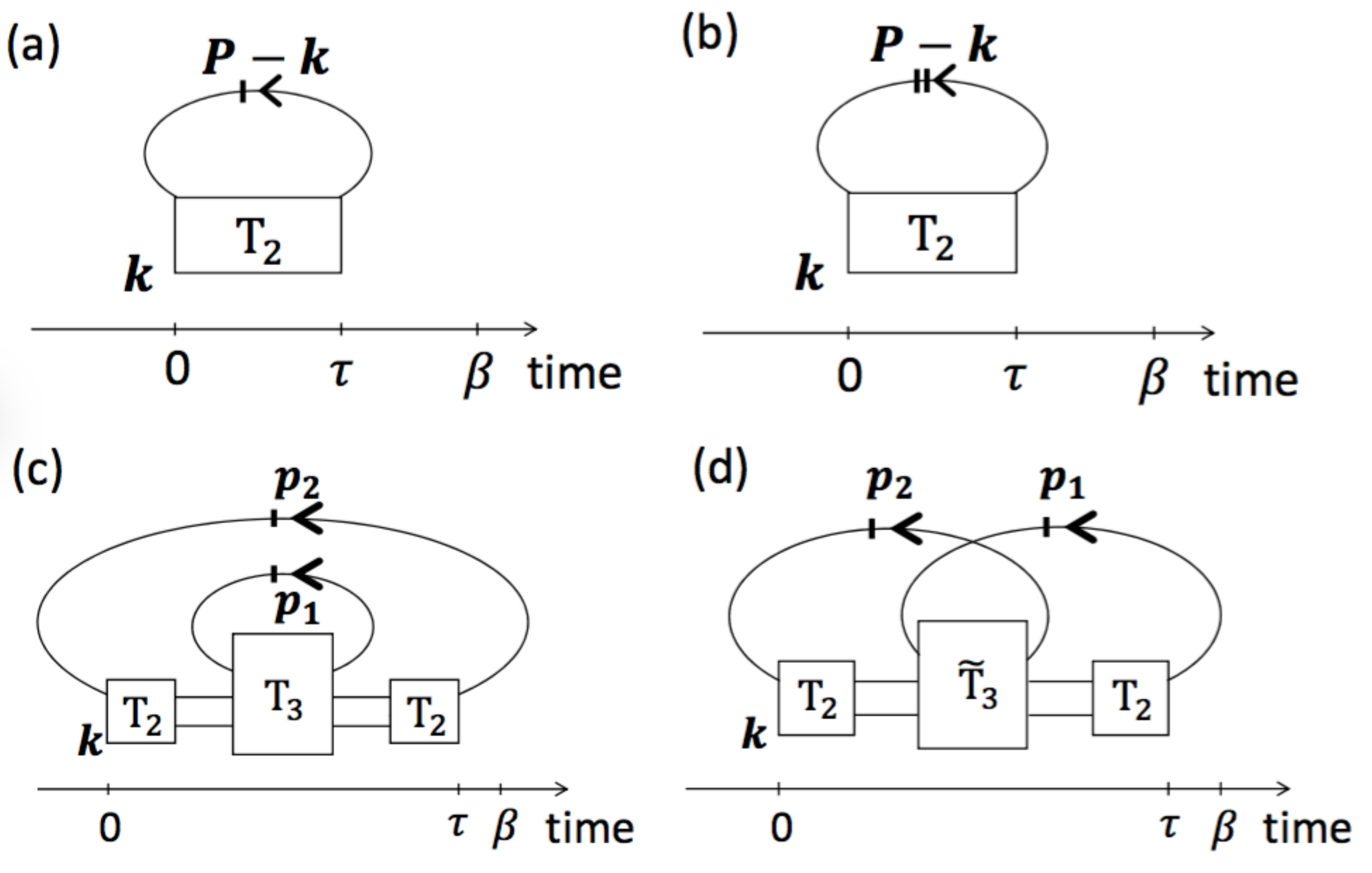}
\caption{Feynman diagrams for the impurity self energy $\Sigma^{(1)}$ (a) and $\Sigma^{(2)}$(b-d). $T_2$ and $T_3$ are respectively the two-body and atom-dimer scattering matrixes; $\tilde{T}_3$ is $T_3$ excluding the first Born term in Eq.\ref{T3} (i.e., excluding reducible diagrams). The boson propagator line with $n$ vertical dashes denotes the $n$-th order contribution $G^{(0,n)}$ in Eq. \ref{G}.}\label{fig2}
\end{figure}

\textit{Formalism.} Here we adopt the diagrammatic approach in the framework of the Virial expansion \cite{Kaplan, Leyronas1,Leyronas2, Hofmann1,Ngampruetikorn,Hofmann2}. The advantage of this method is that it is accurate at high temperature, and can systematically incorporate all the two-body and three-body contributions which allow us to extract the Efimov effect in a controllable way. The Hamiltonian of this $1+N$ system is written as
\begin{equation}
\mathcal{H}=\frac{{\bf p}_{i}^2}{2m_\text{i}}+\sum\limits_{j=1}^{N}\frac{{\bf p}_j^2}{2m_\text{b}}+\sum\limits_{j=1}^{N}V({\bf r}_i-{\bf r}_j),
\end{equation}
where ${\bf r}_i$ and ${\bf p}_i$ label the position and momentum of the impurity atom, while ${\bf r}_j$ and ${\bf p}_j$ ($j=1,\dots,N$) label the position and momentum of $N$ majority bosons. The impurity-boson interaction $V({\bf r})$ is described by an $s$-wave scattering length $a_\text{s}$, which can be tuned across resonance.  Note that here we have neglected the background boson-boson interaction for simplicity. The starting point is to expand the free boson propagator in powers of the fugacity $z_b=e^{\beta\mu_b}$ ($\mu_\text{b}$ is the boson chemical potential, $\beta=1/(k_\text{B}T)$): 
\begin{eqnarray}
G^{(0)}(\mathbf{p},\tau)%&=&e^{-(\epsilon_\mathbf{p}-\mu)\tau} [-\Theta(\tau)-n_b(\epsilon_\mathbf{p}-\mu_b)]\nonumber\\
=e^{\mu\tau}\sum_{n\geq 0} G^{(0,n)}(\mathbf{p},\tau)z_b^n, \label{G}
\end{eqnarray}
where $G^{(0,n)}(\mathbf{p},\tau)$ is $-\Theta(\tau) e^{-\tau\epsilon_\mathbf{p}}$ for $n=0$ and $-e^{-n\beta\epsilon_\mathbf{p}}e^{-\epsilon_\mathbf{p}\tau}$ for $n\geqslant1$; $\tau\in(0,\beta]$ is the imaginary time; $\epsilon_\mathbf{p}={\bf p}^2/(2m_b)$; $n_b(x)=1/(e^{\beta x}-1)$ is the Bose distribution function.
With Eq. \ref{G}, all physical quantities can be expanded in powers of $z_b$. 
 
In Fig. \ref{fig2} we plot the Feynman diagrams for the impurity self-energy $\Sigma({\bf k},\tau)$, which contain all the two-body and three-body diagrams that contribute to the second and the third Virial coefficient ($b_2$ and $b_3$) in the Virial expansion. Fig. \ref{fig2}(a) leads to the lowest order of $\Sigma$ in $z_b$, denoted by
\begin{equation}
\Sigma^{(1)}=z_b \int\frac{d^3\mathbf{P}}{(2\pi)^3} e^{-\beta \epsilon_{\mathbf{P-k}}} T_2\left(\omega+i\delta+\epsilon_{\mathbf{P-k}}-\frac{P^2}{2(m_i+m_b)}\right);
\end{equation}
Fig. \ref{fig2}(b-d) leads to the second order contribution as:
\begin{widetext}
\begin{align}
  & \Sigma^{(2)}=z_b^2 \left\{\int\frac{d^3\mathbf{P}}{(2\pi)^3}  e^{-2\beta \epsilon_{\mathbf{P-k}}}T_2\left(\omega+i\delta+\epsilon_{\mathbf{P-k}}-\frac{P^2}{2(m_i+m_b)}\right)  \right. \nonumber\\
   &+\int\frac{d^3\mathbf{p_1}}{(2\pi)^3}\int\frac{d^3\mathbf{p_2}}{(2\pi)^3} e^{-\beta(\epsilon_{\mathbf{p_1}}+ \epsilon_{\mathbf{p_2}})} T_2^2\left(\omega+i\delta+\Delta-\frac{{p_1'}^2}{2m_{AD}}\right)T_3(\mathbf{p'_1},\mathbf{p'_1},\omega+i\delta+\Delta) \nonumber\\
   &+\left. \int\frac{d^3\mathbf{p_1}}{(2\pi)^3}\int\frac{d^3\mathbf{p_2}}{(2\pi)^3}  e^{-\beta(\epsilon_{\mathbf{p_1}}+ \epsilon_{\mathbf{p_2}})}  T_2\left(\omega+i\delta+\Delta-\frac{{p'_1}^2}{2m_{AD}}\right)\tilde{T}_3(\mathbf{p'_1},\mathbf{p'_2},\omega+i\delta+\Delta) T_2\left(\omega+i\delta+\Delta-\frac{{p'_2}^2}{2m_{AD}}\right) \right\}.
\end{align}
\end{widetext}
Here $\Delta=\epsilon_{\mathbf{p_1}}+\epsilon_{\mathbf{p_2}}-P_t^2/(2M)$, with $\mathbf{P_t}=\mathbf{k}+\mathbf{p_1}+\mathbf{p_2}$ and $M=2m_b+m_i$ respectively the total momentum and the total mass of three-body system; $\mathbf{p'_{1,2}}=\mathbf{p_{1,2}}-m_b\mathbf{P_t}/M$ and $m_{AD}=m_b(m_b+m_i)/M$ are respectively the relative momenta and the reduced mass for atom-dimer scattering. $T_2(E)$ is the two-body scattering matrix with scattering energy $E$:
\begin{equation}
T_2(E)=\frac{2\pi}{m_r} \frac{1}{a_s^{-1}-\sqrt{-2m_r E}}, 
\end{equation}  
where $m_r=\frac{m_bm_i}{m_b+m_i}$ is the reduced mass. $T_3(\mathbf{p_1},\mathbf{p_2},E)$ is the atom-dimer scattering matrix at energy $E$, with $\mathbf{p_1},\mathbf{p_2}$ respectively the relative momenta of the incoming and outgoing atom-dimer states in the center-of-mass frame, and
\begin{eqnarray}
 &&T_3(\mathbf{p_1},\mathbf{p_2},E)=\frac{1}{E-\epsilon_{\cp p_1}-\epsilon_{\cp p_2}-(\cp p_1+\cp p_2)^2/(2m_i)}\nonumber\\
 &&+\int\frac{d^3 \mathbf{q}}{(2\pi)^3}\frac{T_2(E-\frac{q^2}{2m_{AD}})}{E-\epsilon_{\cp p_1}-\epsilon_{\cp q}-\frac{(\cp p_1+\cp q)^2}{2m_i}} T_3(\mathbf{q},\mathbf{p_2},E). \label{T3}
\end{eqnarray}
To this end we have obtained the impurity self-energy, $\Sigma=\Sigma^{(1)}+\Sigma^{(2)}$, up to the order of $z_b^2$. The spectral function can be computed from the propagator of the impurity, $G_i({\cp k},\omega)=(\omega+i\delta-k^2/(2m_i)-\Sigma({\cp k},\omega+i\delta))^{-1}$, as
\begin{equation}
A({\cp k},\omega)=-\frac{1}{\pi} {\rm Im}\Big(G_i({\cp k},\omega)\Big). \label{A}
\end{equation} 

As a benchmark for our calculation, we have obtained the trimer energy $E_T^{(n)}$ at resonance from the pole of $T_3$ and determined the scattering length $a_-^{(n)}$ for the appearance of the $n$-th trimer state in $a_s<0$ side. We have verified that both $a_-^{(n)}$ and $E_T^{(n)}$ well follow the universal scaling law for large $n$, i.e., $a_-^{(n)}/a_-^{(n+1)}=\lambda$, $E_T^{(n)}/E_T^{(n+1)}=\lambda^2$, with $\lambda$ the scaling factor\cite{Efimov,Braaten}. We have also obtained $b_3$ with the same diagrams for $\Sigma^{(2)}$, and the result well reproduces the known analytical behaviors in both unitary and deep molecular regimes \cite{Hofmann2}. 

\textit{Results.} In Table \ref{Table} we compare $\eta$, $\lambda$, $\alpha^{(n)}\equiv 1/(k_Fa_-^{(n)})$ and $\epsilon^{(n)}\equiv E_T^{(n)}/E_F$ for three different impurity-boson(i-b) systems, where $k_F= (6\pi^2 n_b)^{1/3},\ E_F=k_F^2/(2m_b)$ and we take a typical density $n_b=2\times 10^{14}$cm$^{-3}$ for all boson systems. The three-body cutoff is chosen such that the obtained $a_-^{(1)}$ for different systems match the values in Refs.\cite{Aarhus, Greene, scaling_2}. Since the size of the trimer at resonance follows $l_t/d\propto\epsilon^{-1/2}$, the large (or small) $\epsilon$ corresponds to $l_t\ll d$ (or $l_t\gg d$). 
From the table, we can see that the first two systems, $^{39}$K-$^{39}$K(i-b) and $^{40}$K-$^{87}$Rb(i-b), both belong to case (a) in Fig. \ref{schematic}, where the trimers appear only sufficiently close to resonance ($|\alpha^{(n)}|\ll 1$) with their sizes $l_t\gg d$; while the third system, $^{6}$Li-$^{133}$Cs(i-b), belongs to case (b), where the first and the second trimers appear with $|\alpha^{(n)}|\sim 1$, and as varying $1/a_s$, these trimers can have sizes $l_t\sim d$.

\begin{widetext}

\begin{table}[h]%[!hbp]
\begin{tabular}{|c|c|c|c|c|c|c|c|c|}
  \hline
 impurity-boson   & $\eta$ & $\lambda$ & $\alpha^{(1)}$ & $\alpha^{(2)}$ & $\alpha^{(3)}$ & $\epsilon^{(1)}$ & $\epsilon^{(2)}$ & $\epsilon^{(3)}$  \\
  \hline
  $^{39}$K-$^{39}$K \cite{Aarhus}& 1 & 1986 & $-2.76\times 10^{-3}$ & $-1.39\times 10^{-6}$ & $-6.99\times 10^{-10}$ & $1.51\times 10^{-4}$  & $3.84\times 10^{-11}$ & $9.73\times 10^{-18}$ \\
  \hline
  $^{40}$K-$^{87}$Rb \cite{JILA} & $2.2$ & 123 & $-2.76\times 10^{-2}$ & $-2.23\times 10^{-4}$ & $-1.82\times 10^{-6}$  & $1. 40\times 10^{-2}$& $9.32\times 10^{-7}$ &  $6.19\times 10^{-11}$ \\
  \hline
  $^{6}$Li-$^{133}$Cs \cite{scaling_2,scaling_3} & $22.2$ & $4.87$ & $-2.56$ & $-0.40$ & $-7.87\times 10^{-2}$  & 185.9  &  6.09 & 0.25  \\
  \hline
\end{tabular}
\caption{Mass ratio ($\eta\equiv m_b/m_i$), Efimov scaling factor from zero-range theory ($\lambda$), the interaction parameter for the appearance of the $n$-th Efimov trimers ($\alpha^{(n)}\equiv 1/(k_F a_-^{(n)})$), and the $n$-th trimer energy at resonance ($\epsilon^{(n)}\equiv E_T^{(n)}/E_F$) for three systems with different impurity-boson(i-b) combinations. The values of $a_-^{(1)}$ are from Refs.\cite{Aarhus, Greene, scaling_2}.  For Li-Cs system, the relatively large derivation of the scaling factor for the two lowest trimer states from $\lambda$ is due to the very deep lowest trimer (with $E_T^{(1)}$ of the order of the cutoff energy), in which the finite range effect becomes non-negligible.
\label{Table}
}
\end{table}

\end{widetext}

\begin{figure}[t] 
\includegraphics[height=10cm,width=8cm]{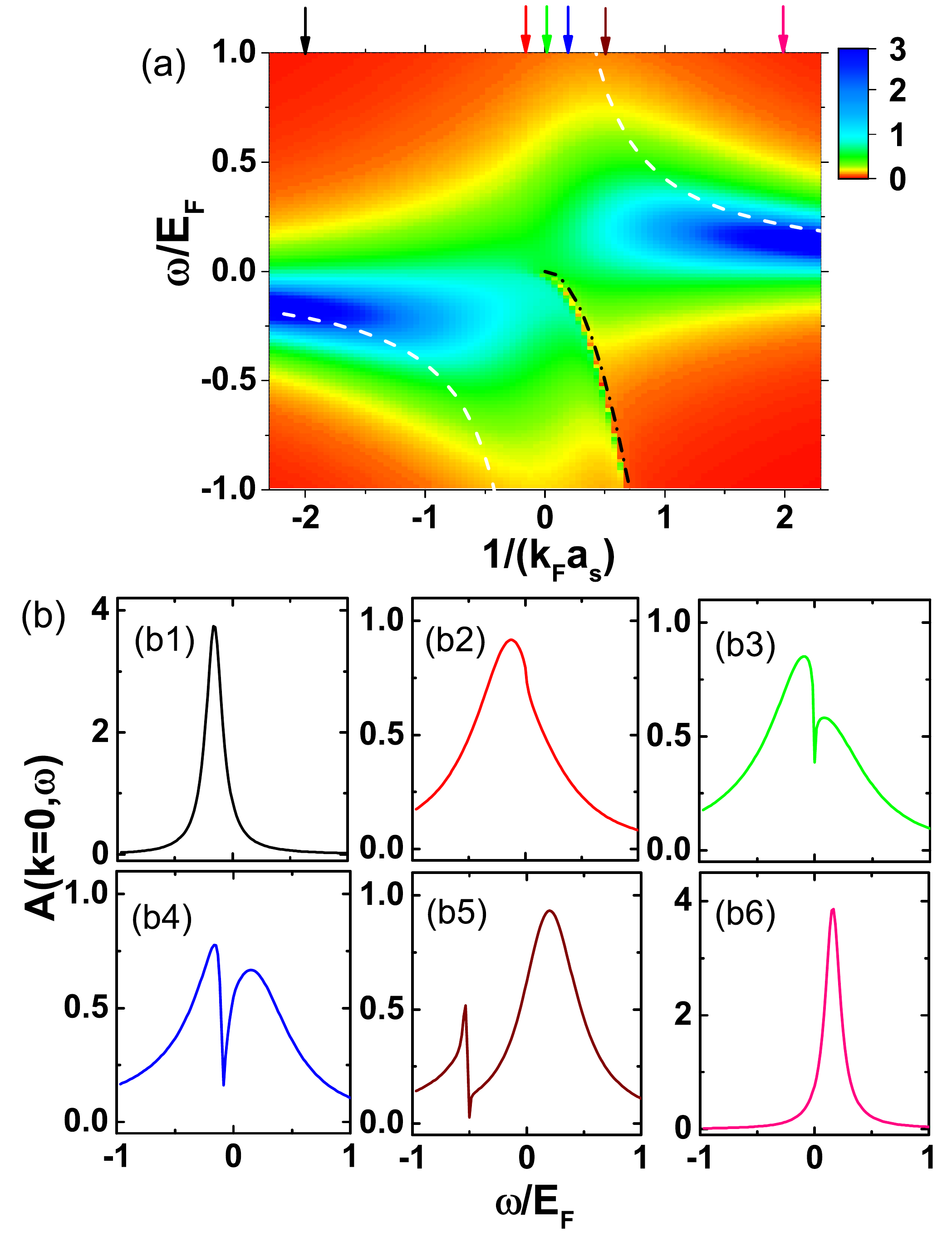}
\caption{(Color online). Spectral function $A(\cp k=0, \omega)$ (in unit of $1/E_F$) for $^{39}$K-$^{39}$K(i-b) system. (a) shows the contour plot of $A(\cp k=0, \omega)$ in term of $1/(k_Fa_s)$ and $\omega/E_F$. For comparison, we show the two-body binding energy (black dashed-dot) and the mean-field energies of attractive and repulsive branches (white dashed). (b) shows slices of $A(\cp k=0, \omega)$ for $1/(k_Fa_s)=-2, -0.2, 0, 0.2, 0.5, 2$ (from (b1) to (b6)), as labeled by the arrows in (a) with according colors. Here $z_b=0.1$. }\label{fig3}
\end{figure}

\begin{figure}[t] 
\includegraphics[height=10cm,width=8cm]{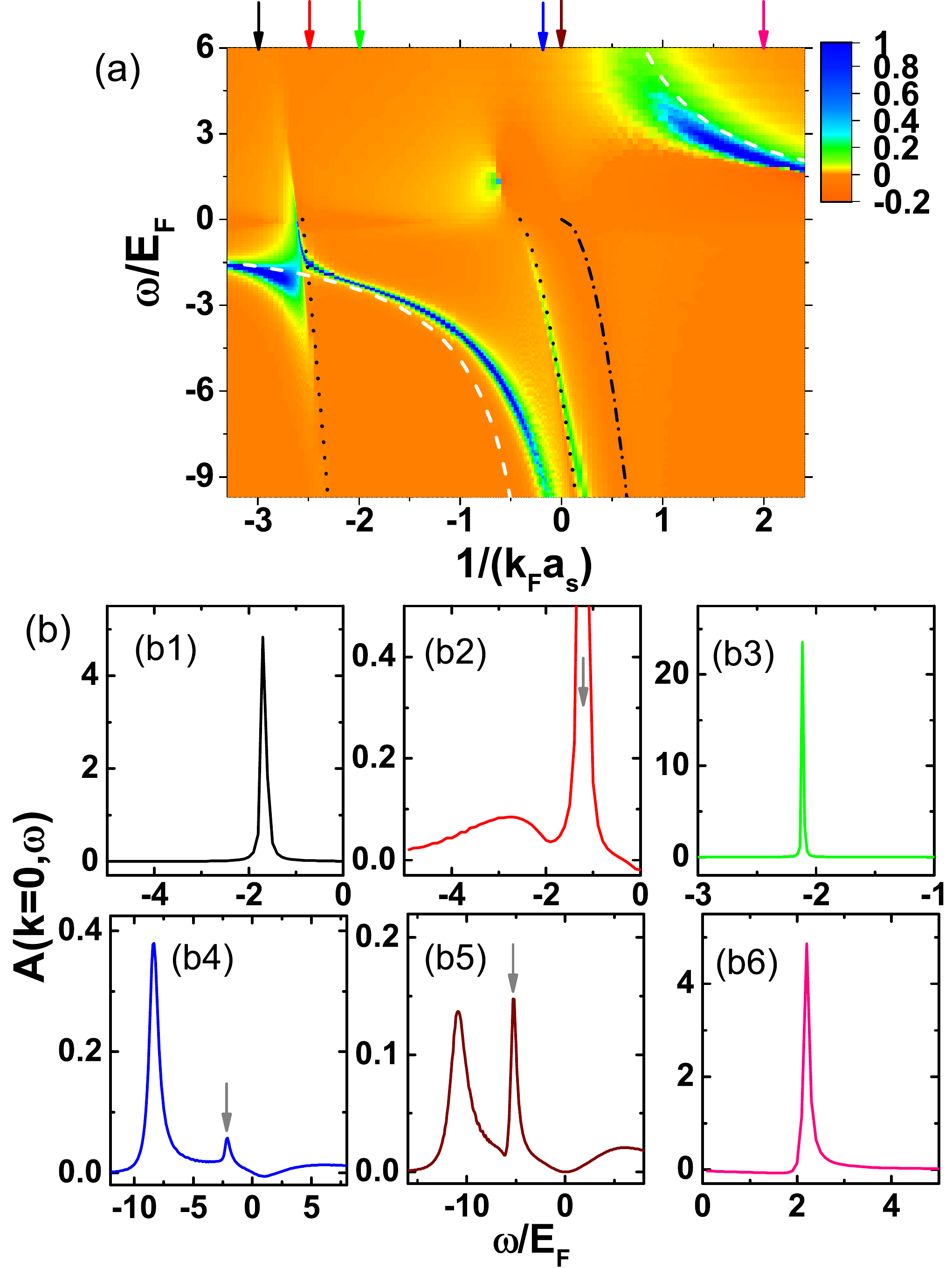}
\caption{(Color online). Same plot of $A(\cp k=0, \omega)$ as in Fig. \ref{fig3} but for $^{6}$Li-$^{133}$Cs (i-b) system. From (b1) to (b6), $1/(k_Fa_s)=-3, -2.5, -2, -0.2, 0, 2$. The additional dotted lines in (a) show the energies of the first and the second Efimov trimers from three-body calculations. The gray arrows in (b2,b4,b5) mark the spectral peaks of the Efimov branches. }\label{fig4}
\end{figure}

Below we present the spectral results for $^{39}$K-$^{39}$K(i-b) system (Fig. \ref{fig3}) and  $^{6}$Li-$^{133}$Cs(i-b) system (Fig. \ref{fig4}) as the representatives of two cases in Fig. \ref{schematic}. Since the injection spectroscopy used in the experiments \cite{Grimm,Kohl,Grimm2016,Roati,Aarhus,JILA} can be described by $A(\cp k=0,\omega)$, taking $z_b=0.1$ for both systems (giving the thermal wavelength $\lambda_{T}=0.47 d$), we show the contour plots of $A(0,\omega)$ in terms of $1/(k_F a_s)$ and $\omega/E_F$ in Fig. \ref{fig3}(a) and \ref{fig4}(a), and slices of $A(0,\omega)$ in Fig. \ref{fig3}(b) and \ref{fig4}(b) for several typical values of $-1/(k_F a_s)$ across resonance. 

In Fig. \ref{fig3}, the spectrum shows only attractive and repulsive polaron branches without any signature of Efimov physics. 
It shows that, at least in the temperature regime we are considering, the Efimov signature is not visible if $l_t\gg d$. 
As increasing $1/(k_Fa_s)$, the spectrum starts from a well-defined quasi-particle peak centered at negative $\omega$ (attractive polaron, Fig. \ref{fig3}(b1)), which gradually becomes broader (Fig. \ref{fig3}(b2)) to exhibit a two-peak structure near resonance (Fig. \ref{fig3}(b3-b5), and finally evolves to a single peak centered at positive $\omega$ (repulsive polaron, Fig. \ref{fig3}(b6)). All these features are consistent with current experimental observations \cite{Aarhus, JILA}.

Contrarily, in Fig. \ref{fig4}, besides the attractive and repulsive branches, there are two visible Efimov branches, which are associated with the first two Efimov trimers in vacuum emerging at $a_-^{(1,2)}<0$ (dotted lines in Fig. \ref{fig4}(a)). Interestingly, these Efimov branches can be very close or even level-crossing with the  attractive branch of polarons, and the inter-branch hybridization leads to a much broadened spectrum near $a_s\sim a_-^{(1)}$ as well as an enhanced signal of the second Efimov branch near resonance\cite{footnote}.

In this case, starting from a single attractive polaron branch (Fig. \ref{fig4}(b1)), with the increase of $1/(k_Fa_s)$, one can see the first Efimov branch appears around $a_s\sim a_-^{(1)}$ with a narrow peak near zero frequency (marked by the arrow in Fig. \ref{fig4}(b2)) and it hybridizes with the attractive polaron branch. %Furthering increasing $1/(k_Fa_s)$, the first Efimov state quickly becomes very deep and becomes invisible when $l_t\ll d$, and there is again a single attractive polaron branch visible (Fig. \ref{fig4}(b3)). 
This Efimov branch quickly merges into the attractive polaron branch away from their (avoided) level-crossing (Fig. \ref{fig4}(b3)). %Until the second Efimov branch appears nearby resonance (Fig. \ref{fig4}(b4)), two branches co-exist at negative frequency (Fig. \ref{fig4}(b4,b5)). 
The second Efimov branch shows up as a nearby resonance (arrow in Fig. \ref{fig4}(b4)), and its signal can become more pronounced when its level moves closer to the attractive branch (Fig. \ref{fig4}(b5)). Finally it becomes a single branch at positive frequency (repulsive polaron, Fig. \ref{fig4}(b6)).

Note that the Efimov branches shown in Fig. \ref{fig4} are only visible after including the three-body contributions ($\Sigma^{(2)}$). In contrast, we have checked that the inclusion of $\Sigma^{(2)}$ in Fig. \ref{fig3} does not make  qualitative change to the spectrum. This confirms the distinct roles of three-body effect played in the two systems, as illustrated in Fig. \ref{schematic}. 

%In fact, the other shallow Efimov trimers near resonance, though not visible in Fig.3(b1), also play essential roles. These states, with all effects incorporated in $\Sigma^{(2)}$, can greatly reduce the spectral signal built by two-body correlations (only $\Sigma^{(1)}$) and lead to nearly disconnected attractive and repulsive branches in Fig.3(b1). This is in sharp contrast to the resonance regime of case (a) as shown in Fig.3(a1). 

%\begin{figure}[t] 
%\includegraphics[width=0.5\textwidth]{fig5.pdf}
%\caption{(Color online). Relative spectral width, $\sigma/\bar{\omega}$, as a function of $1/(k_Fa_s)$ for the $^{39}$K-$^{39}$K(i-b) spectrum in Fig. \ref{fig3} and $^{6}$Li-$^{133}$Cs (i-b) spectrum in Fig. \ref{fig4).  $\sigma$ and $\bar{\omega}$ are extracted from a Gaussian fit of each distinguishable peak in the spectrum: $A()$ }\label{fig5}
%\end{figure}

Another notable difference between Fig. \ref{fig3} and Fig. \ref{fig4} is that in the latter, the attractive and repulsive branches have much narrower relative spectral width, defined by the ratio of the absolute width to the mean location of the spectral peak. 
%$\Delta \omega/\bar{\omega}$, where $\Delta \omega$ and $\bar{\omega}$ are respectively the width and the mean location of the spectral peak. 
Near resonance, these branches are well separated and disconnected, unlike those in Fig. \ref{fig3}. This suggests that for given $\lambda_T/d$, the Bose polaron quasi-particle is more well-defined for larger mass ratio $\eta$. 
%This can be attributed to the large inertia of the light impurities, whose status can be relatively harde combined effects of two-body correlation modified by the large mass imbalance and an infinite number of dense Efimov trimers near resonance. 
%All these features can be directly detected in Li-Cs atomic gases or other Bose polarons with large mass imbalance.  

\textit{Discussion and Outlook.} 
In this work, we have revealed the signature of Efimov physics in the spectral response of the Bose polarons with large mass imbalance. The setting here is different from previous ones exploring the energetics of the Bose polarons with relatively small mass imbalance\cite{Levinsen2, Giorgini}. Nevertheless, the phenomenon of avoided level crossing shown in Fig.\ref{fig4} is physically in accordance with atom-trimer continuity in the ground state of the Bose polarons as studied in Ref.\cite{Levinsen2}.

%Our work demonstrates two essential conditions for the observation of Efimov physics in the spectral response of Bose polarons, i.e., the Efimov trimers have sizes comparable to mean distance of many-body background( $l_t\sim d$), and they have considerable overlap with the background (requiring $|k_Fa_s|\sim 1$). The first condition rules out both very shallow and very deep Efimov trimers, while the second condition rules out trimers far from resonance. These conditions should generally apply to a wide class of polaron systems, even including Fermi polarons with large mass imbalance where Efimov physics is possible\cite{Petrov}. 

Our results (assuming no interaction between bosons) can be directly probed in $^{6}$Li-$^{133}$Cs atomic system near $B_0=889$G Feshbach resonance\cite{new_expt1,new_expt2}, where the boson-boson scattering length ($a_{bb}$) is small and the Efimov scenario is not modified by finite $a_{bb}$ (except for the ground state trimer\cite{new_expt1}). Moreover, the diagrammatic approach we used in this work can be generalized to interacting boson systems, with the boson-boson interaction contributing to another scattering channel. Our method can also be systematically improved to include $n$-body correlations ($n>3$) in a controllable matter, for instance, the effect of four-body bound states consisting of one $^{6}$Li and three $^{133}$Cs atoms \cite{Blume2}, which may result in additional signals near the location of their appearances. 

In principle, our results can also be applied to the Fermi polarons. However, for the reduced three-body problem from Fermi polarons, the Efimov states appear only when the mass ratio exceeds $13.6$\cite{Petrov}. Just above the critical mass ratio, the Efimov trimers are shallow and the scaling factor is large, so the system falls into the scenario (a) discussed here. Thus, in order to observe the signature of Efimov physics in Fermi polarons, one needs the mass ratio far exceeding $13.6$. 

{\it Acknowledgements.} We thank Ran Qi, Cheng Chin, Jesper Levinsen, Ren Zhang, Pengfei Zhang and Zhigang Wu for helpful discussions, and the Supercomputer Center in Guangzhou for computational support. 
%We also thank the Supercomputer Center in Guangzhou for support. 
This work is supported by the National Natural Science Foundation of China (No. 11626436, 11374177, 11421092, 11534014, 11325418), the National Key Research and Development Program of China (No. 2016YFA0300603, 2016YFA0301600), and Tsinghua University Initiative Scientific Research Program.

%\begin{widetext}

%\vspace{0.05in}

%\section{Supplementary Material}

%In this supplementary material we present some details of 
%solving two-body and three-body problem in the presence of spin-orbit coupling considered in the main text.

%Include:

%1) evaluation of Eq.2-6 and their associated Feynman diagrams;

%2) few-body check: $a_-, E_T, b3$

%3) how to obtain table I: parameters chosen, cutoff in numerics

%4) if only take T2 contribution, how Fig.3(b1) looks like

%\end{widetext}


\begin{thebibliography}{99}
\bibitem{footnote} In Fig. \ref{fig4} there exists a very small parameter regime nearby zero energy where $A(0,\omega)$ becomes negative after including $\Sigma^{(2)}$, and we attribute this to the convergency problem of the Virial expansion and the absence of higher-order contributions of $\Sigma$ beyond the $z_b^2$ level. %In Fig. \ref{fig4} there are some parameter region with negative $A(0,\omega)$, which can be attributed to the higher-order contributions of $\Sigma$ beyond the $z_b^2$ level in this work.
\end{thebibliography}
\end{document}